\documentclass[a4paper,11pt]{article}

\usepackage{jheppub} 

\usepackage[T1]{fontenc} 

\usepackage{threeparttable}
\usepackage{multirow}
\usepackage{multicol}
\usepackage{subfigure}
\usepackage{amssymb,bm,mathrsfs,bbm,amscd}
\usepackage{float}
\usepackage{units}
\usepackage{lineno}

\def \ee   {e^{+}e^{-}}
\def \piz  {\pi^{0}}
\def \pip  {\pi^{+}}
\def \pim  {\pi^{-}}

\def \ipb  {\mbox{pb$^{-1}$}}
\def \ifb  {\mbox{fb$^{-1}$}}
\def \gev  {\mbox{GeV}}
\def \gevc {\mbox{GeV/$c$}}
\def \gevcc{\mbox{GeV/$c^2$}}

\def \ebeam {E_{\mathrm{beam}}}

\def \Ks {K^0_{S}}
\def \KL {K^0_{L}}

\def \Modea {\bar{p}K_{S}^0}
\def \Modeb {\bar{p}K^{+}\pi^-}
\def \Modec {\bar{p}K_{S}^0\pi^0}
\def \Moded {\bar{p}K_{S}^0\pi^-\pi^+}
\def \Modee {\bar{p}K^{+}\pi^-\pi^0}
\def \Modef {\bar{p}\pi^-\pi^+}
\def \Modeaa {\bar{\Lambda}\pi^-}
\def \Modebb {\bar{\Lambda}\pi^-\pi^0}
\def \Modecc {\bar{\Lambda}\pi^-\pi^+\pi^-}
\def \Modeaaa {\bar{\Sigma}^{0}\pi^-}
\def \Modebbb {\bar{\Sigma}^{-}\pi^0}
\def \Modeccc {\bar{\Sigma}^{-}\pi^-\pi^+}

\def \lcp {\Lambda_{c}^{+}}
\def \lcm {\bar{\Lambda}_{c}^{-}}

\def \NST {N^{\mathrm{ST}}}
\def \NDT {N^{\mathrm{DT}}}

\def \effST {\varepsilon^{\mathrm{ST}}}
\def \effDT {\varepsilon^{\mathrm{DT}}}

\def \MMsq {M^2_{\mathrm{miss}}}

\newcommand{\BR}{\mathcal{B}}

\newcommand{\jpsi}{J/\psi}

\arxivnumber{2406.18083} 

\title{\boldmath Measurements of $K_S^0$-$K_L^0$ asymmetries in the decays $\Lambda_c^+ \to pK_{L,S}^0$, $pK_{L,S}^0\pi^+\pi^-$ and $pK_{L,S}^0\pi^0$} 

\collaborationImg{\includegraphics[width=0.15\textwidth, angle=90]{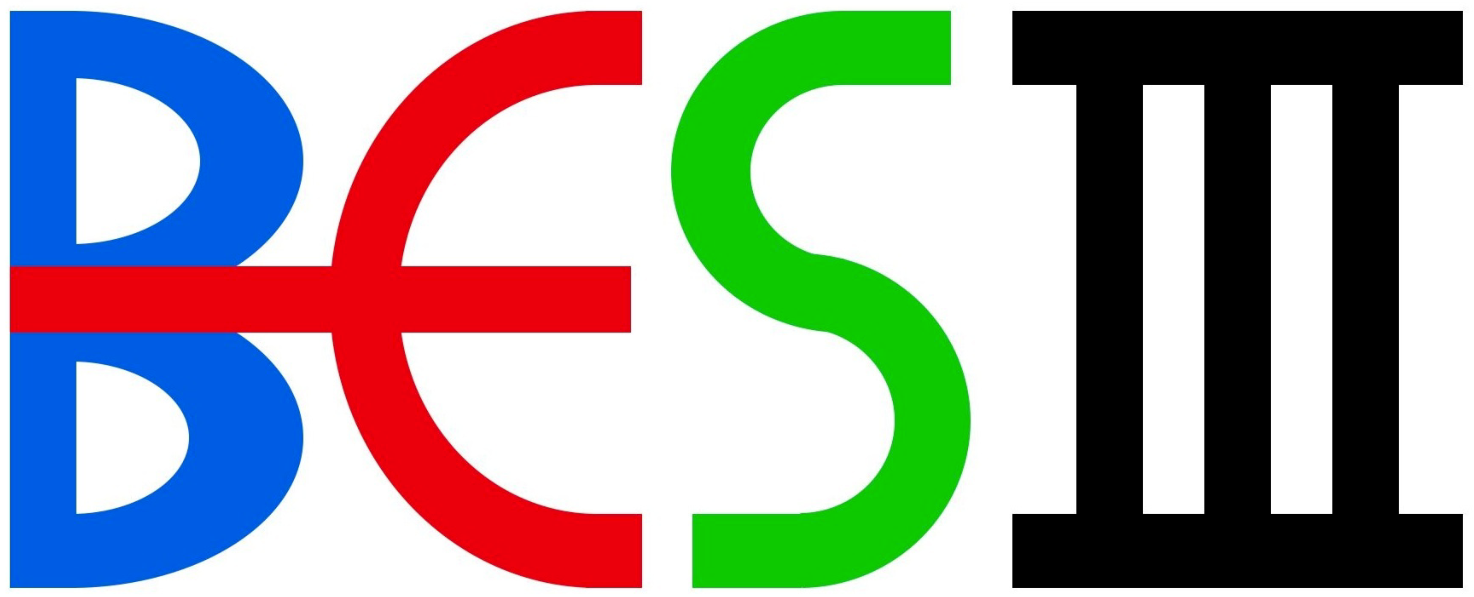}}
\collaboration{The BESIII collaboration}
\emailAdd{besiii-publications@ihep.ac.cn}

\abstract{ Using $e^+e^-$ annihilation data sets corresponding to an
  integrated luminosity of 4.5 $\text{fb}^{-1}$, collected with the BESIII
  detector at center-of-mass energies between 4.600 and 4.699 GeV,
  we report the first measurements of the absolute branching fractions
  $\mathcal{B}(\Lambda_c^+\to pK_{L}^{0})=(1.67 \pm 0.06 \pm 0. 04)\%$, $\mathcal{B}(\Lambda_c^+\to
  pK_{L}^{0}\pi^+\pi^-)=(1.69 \pm 0.10 \pm 0.05)\%$, and $\mathcal{B}(\Lambda_c^+\to
  pK_{L}^{0}\pi^0)=(2.02 \pm 0.13 \pm 0.05)\%$, where the first uncertainties
  are statistical and the second systematic.  Combining with the known
  branching fractions of $\Lambda_c^+ \to pK_{S}^{0}$, $\Lambda_c^+ \to pK_{S}^{0}\pi^+\pi^-$, and
  $\Lambda_c^+ \to pK_{S}^{0}\pi^0$, we present the first measurements of the
  $K_{S}^{0}$-$K_{L}^{0}$ asymmetries $R(\Lambda_c^+, K_{S,L}^0X) = \frac{\mathcal{B}(\Lambda_c^+ \to K_{S}^{0} X) - \mathcal{B}(\Lambda_c^+ \to K_{L}^{0} X)}{\mathcal{B}(\Lambda_c^+ \to K_{S}^{0} X) + \mathcal{B}(\Lambda_c^+ \to K_{L}^{0} X)}$ in charmed baryon decays: $R(\Lambda_c^+, pK_{S,L}^0) = -0.025 \pm 0.031$, $R(\Lambda_c^+, pK_{S,L}^0\pi^+\pi^-) = -0.027 \pm 0.048$, and
  $R(\Lambda_c^+, pK_{S,L}^0\pi^0) =-0.015 \pm 0.046$. No significant
  asymmetries with statistical significance are observed.
}


\begin{document}
\maketitle
\flushbottom

\section{Introduction}
\label{sec:intro}
\hspace{1.5em}

The lightest charmed baryon, $\lcp$, provides a unique environment for
studying the behavior of light di-quarks in the presence of a heavy
quark~\cite{Cheng:2021qpd}. Its hadronic decays occur only through the
weak interaction, and various theoretical models have been
proposed. These include the covariant confined quark
model~\cite{Korner:1992wi,Ivanov:1997ra}, the pole
model~\cite{Cheng:1991sn, Xu:1992vc,
  Cheng:1993gf,Xu:1992sw,Zenczykowski:1993jm,Sharma:1998rd}, current
algebra~\cite{Korner:1978ec,Uppal:1994pt}, and SU(3) flavor symmetry
approaches~\cite{Lu:2016ogy,Savage:1989qr,Kohara:1991ug,Verma:1995dk,Chau:1995gk}. Its
decays fall into three categories: Cabibbo-favored (CF) decays, singly
Cabibbo-suppressed decays, and doubly Cabibbo-suppressed (DCS) decays.
The decay amplitudes of the CF and DCS modes are expected to be
proportional to the products of the Cabibbo-Kobayashi-Maskawa elements
$|V_{ud}^*V_{cs}|$ and $|V_{us}^*V_{cd}|$, respectively. The ratio of
their decays is approximately of the order of $\mathcal{O}(10^{-3})$,
resulting in a small branching fraction (BF) for the DCS decay and
making it challenging to observe directly in experiments.

In addition to direct measurements of DCS decays, the amplitudes of
DCS modes can be probed using the $\Ks$-$\KL$ asymmetry in the decays
into neutral kaons, which arises from the interference
between CF and DCS amplitudes~\cite{Bigi:1994aw,Wang:2017gxe}. The
$\Ks$-$\KL$ asymmetry has been studied in the decays of charmed $D$
mesons, where the asymmetry is defined as
\begin{equation}\label{eq:dmeson_asymmetry}
    R(D, K_{S,L}^0X) = \frac{\BR(D\to\Ks X) - \BR(D\to\KL X)}{\BR(D\to\Ks X)+\BR(D\to \KL X)},
\end{equation}
and $X$ can be
$\piz$, $\eta$, $\eta^\prime$, $\omega$, $\rho^0$ or $\phi$.  
A large asymmetry of $R(D^0, K_{S,L}^0\piz) = 0.108\pm0.025\pm0.024$~\cite{CLEO:2007rhw} was reported in a previous measurement by the CLEO experiment, where the first uncertainty is statistical and the second is systematic.
The BESIII experiment reported measurements of the $\Ks$-$\KL$ asymmetries
$R(D^0, K_{S,L}^0X)$, where $X=\phi$, $\eta$,
$\omega$~\cite{BESIII:2022xhe}. Significant asymmetries were observed
in $D^0\to\KL\eta$ and $D^0 \to \KL\eta^\prime$ decays with
$R(D^0,K_{S,L}^0\eta) = 0.080\pm0.022$ and
$R(D^0,K_{S,L}^0\eta^\prime)=0.108\pm0.035$, respectively.  In
addition, this asymmetry has been investigated for the lightest
charmed strange meson, and $R(D_s^+, K_{S,L}^0K^+)$ was determined to
be $(-2.1\pm1.9\pm1.6)\%$~\cite{BESIII:2019kfh}. However, such
measurements have not been made for the decays of charmed
baryons.

Using flavor SU(3)
asymmetry~\cite{Lu:2016ogy,Savage:1989qr,Kohara:1991ug,Verma:1995dk,Chau:1995gk},
theoretical predictions~\cite{Wang:2017gxe} for $\Ks$-$\KL$
asymmetries have been made for charmed baryon two-body decays into a light
baryon and a neutral kaon. Similar to
Equation~\ref{eq:dmeson_asymmetry}, the asymmetry of $\BR(\lcp\to \Ks
X)$ and $\BR(\lcp \to \KL X)$ in charmed baryon decays is defined as
\begin{eqnarray}
    R(\lcp, K_{S,L}^0 X)=\frac{\BR(\lcp\to \Ks X)-\BR(\lcp\to \KL X)}{\BR(\lcp\to \Ks X)+\BR(\lcp\to \KL X)},
    \label{eq:asymmetry}
\end{eqnarray}
where $X$ is $p$, $p\pip\pim$ or $p\piz$. Equation~\ref{eq:asymmetry}
can be further reduced as $R(\lcp \to K^0_{S,L}X) \simeq
-2r_{f}\cos{\delta_f}$, where $r_f$ and $\delta_f$ are the relative
strength and phase between the DCS ($\lcp \to K^0X$) and CF ($\lcp \to
\bar{K}^0X$) amplitudes, respectively. The parameter $r_f$ is expected
to be proportional to the ratio $|V_{cd}^*V_{us}/V_{cs}^*V_{ud}| \sim
\lambda^2 $~\cite{ParticleDataGroup:2022pth}. A non-zero asymmetry
value indicates the presence of DCS processes. The asymmetry of $\lcp
\to pK_{S,L}^0$ is predicted to be in the range of ($-0.010$, $0.087$)
in Ref.~\cite{Wang:2017gxe}. The $\Ks$-$\KL$
asymmetry is a promising observable with which to search
for the two-body DCS processes of charmed baryons.

In this paper, we report the first measurements of the absolute BFs of
$\lcp \to p\KL$, $\lcp\to p\KL\pip\pim$ and $\lcp\to p\KL\piz$ based
on $\ee$ annihilation data samples corresponding to a total integrated luminosity
of 4.5 $\ifb$ collected at the center-of-mass (c.m.) energies
$\sqrt{s}$ between 4.600 and 4.699 GeV. The luminosities are listed in
Table~\ref{tab:data_samples}~\cite{BESIII:2022dxl,BESIII:2022ulv}. Using
the results of $\BR(\lcp\to p\Ks)$, $\BR(\lcp\to p\Ks\pip\pim)$, and
$\BR(\lcp\to p\Ks\piz)$ from the Particle Data Group
(PDG)~\cite{ParticleDataGroup:2022pth}, we present the $\Ks$-$\KL$
asymmetries $R(\lcp, K_{S,L}^0 X)$, where $X = p$, $p\pip\pim$ or
$p\piz$. Charge conjugate channels are implied throughout this paper,
unless explicitly stated.

\begin{table}[!hpt]
	\begin{center}
\caption{The integrated luminosities at each c.m. energy~\cite{BESIII:2022dxl,BESIII:2022ulv}.}
\label{tab:data_samples}
\begin{tabular}{c|c}
\hline \hline
$\sqrt{s}$ (GeV) & Integrated luminosity $(\ipb)$ \\\hline
4.600 & 586.9 $\pm$ 0.1 $\pm$ 3.9 \\
4.612 & 103.7 $\pm$ 0.1 $\pm$ 0.6 \\
4.628 & 521.5 $\pm$ 0.1 $\pm$ 2.8 \\
4.641 & 551.7 $\pm$ 0.1 $\pm$ 2.9 \\ 
4.661 & 529.4 $\pm$ 0.1 $\pm$ 2.8 \\ 
4.682 & 1667.4 $\pm$ 0.2 $\pm$ 8.8 \\
4.699 & 535.5 $\pm$ 0.1 $\pm$ 2.8 \\
\hline \hline
\end{tabular}	
\end{center}
\end{table}

\section{BESIII experiment and Monte Carlo simulation}
\label{sec:detector}
\hspace{1.5em} The $\uchyph=0$BESIII detector~\cite{BESIII:2009fln}
records symmetric $\ee$ collisions provided by the $\uchyph=0$BEPCII
storage ring~\cite{Yu:2016cof}, which operates at c.m. energies
ranging from 1.85 to 4.95 GeV, with a peak luminosity of $1.1 \times
10^{33}\;\mathrm{cm}^{-2}\mathrm{s}^{-1}$ achieved at $\sqrt{s} =
3.773\;\mathrm{GeV}$. The $\uchyph=0$BESIII detector has collected
large data samples in this energy region~\cite{BESIII:2020nme}.  The
cylindrical core of the $\uchyph=0$BESIII detector covers 93\% of the
full solid angle and consists of a helium-based multilayer drift
chamber (MDC), a plastic scintillator time-of-flight system (TOF), and
a CsI(Tl) electromagnetic calorimeter (EMC), which are all enclosed in
a superconducting solenoidal magnet providing a 1.0 T magnetic
field~\cite{Huang:2022wuo}. The solenoid is supported by an octagonal
flux-return yoke with resistive plate counter muon identification
modules interleaved with steel.  The charged-particle momentum
resolution at $1\,\gevc$ is $0.5\%$, and the $\mathrm{d}E/\mathrm{d}x$
resolution is $6\%$ for electrons from Bhabha scattering. The EMC
measures photon energies with a resolution of $2.5\%$ ($5\%$) at
$1\,\gev$ in the barrel (end cap) region.  The time resolution in the
TOF barrel region is 68 ps, while that in the end cap region was
initially 110 ps. The end cap TOF system was upgraded in 2015 using
multi-gap resistive plate chamber technology, providing a time
resolution of 60 ps~\cite{Li:2017jpg,Guo:2017sjt,Lange:2001uf}. Of the
data used in this analysis, 87\% was with the upgraded end cap TOF.

Simulated samples generated with
\textsc{geant4}-based~\cite{GEANT4:2002zbu} Monte Carlo (MC) software,
which includes the geometric description of the BESIII detector and
the detector response
performance~\cite{YouZhengYun2008,Liang:2009zzb,Huang:2022wuo}, are
used to determine detection efficiencies and to estimate potential
background contributions. The simulation describes the beam energy
spread and the initial state radiation (ISR) in the $e^+e^-$
annihilations with the generator
\textsc{kkmc}~\cite{Jadach:2000ir,Jadach:1999vf}. The inclusive MC
samples, corresponding to about 40 times the number of events of the
data samples, include the production of $\lcp\lcm$ pairs, open charm
processes, the ISR production of vector charmonium(-like) states, and
the continuum processes incorporated in
\textsc{kkmc}~\cite{Jadach:2000ir,Jadach:1999vf}. The known decay
modes are modeled with \textsc{evtgen}~\cite{Lange:2001uf,Ping:2008zz}
using BFs taken from the PDG~\cite{ParticleDataGroup:2022pth}, and the
remaining unknown charmonium decays are modeled with
\textsc{lundcharm}~\cite{Chen:2000tv,Lange:2001uf}. Final state
radiation from charged final state particles is incorporated using
\textsc{photos}~\cite{Richter-Was:1992hxq}. For the production of
$e^+e^- \to \lcp\lcm$ events, the Born cross-section line shape from
BESIII measurements is used~\cite{BESIII:2017kqg,BESIII:2023rwv}.
Exclusive $\ee \to \lcp\lcm$ signal MC samples are generated with 
$\lcp$ decaying to $p\KL$, $p\KL\pip\pim$ and $p\KL\piz$ in half of 
the signal events, and $\lcm$ decaying to the signal final states for 
the other half. The remaining $\lcm$ and $\lcp$ are required to decay 
to twelve specific tag modes, as detailed in Section~\ref{sec:ana}.
The angular distribution of the decay $\lcp \to p\KL$
is modeled with decay asymmetry parameters obtained from
Ref.~\cite{BESIII:2019odb}. For processes from $\lcp \to p\KL\pip\pim$
and $\lcp \to p\KL\piz$ channels, signal models are tunned based on the data.
Additional MC samples are generated to estimate contributions
from peaking background processes, where $\lcm$ decays into tag modes
and $\lcp$ decays into $p\Ks$, $p\eta$, $p\Ks\piz$, and
$p\Ks\pip\pim$, with $\Ks$ and $\eta$ decaying inclusively. Each tag
mode of the exclusive MC samples is generated with the same number of
events.

\section{Data analysis}
\label{sec:ana}
\hspace{1.5em} Taking advantage of the threshold production of the
$\lcp\lcm$ pair, the double-tag (DT) method~\cite{MARK-III:1985hbd,
MARK-III:1987jsm,Ke:2023qzc,Li:2021iwf} is employed to study $\lcp\to p\KL$,
$\lcp\to p\KL\pip\pim$ and $\lcp\to p\KL\piz$, where $\KL$ is
reconstructed by the missing-mass technique. A single-tag (ST) event
is selected by tagging a $\lcm$ baryon with one of the following
twelve tag modes: $\Modea$, $\Modeb$, $\Modec$, $\Moded$, $\Modee$,
$\Modef$, $\Modeaa$, $\Modebb$, $\Modecc$, $\Modeaaa$, $\Modebbb$, and
$\Modeccc$. The ST event selection criteria, efficiencies, and yields
are described in Ref.~\cite{BESIII:2022xne}. The signal decays
$\lcp\to p\KL$, $p\KL\pip\pim$, and $p\KL\piz$ are reconstructed using
the remaining charged tracks and photons recoiling against the ST
$\lcm$ candidates, and referred to as DT events.

Charged tracks are required to be within $|\cos\theta|<0.93$, where
$\theta$ is the polar angle defined with respect to the $z$-axis,
which is the symmetry axis of the MDC. The distance of closest
approach to the interaction point (IP) must be less than 10 cm along
the $z$ axis and less than 1 cm in the perpendicular plane. Particle
identification (PID) for charged tracks combines measurements of the
energy deposited in the MDC (d$E/$d$x$) and the flight time in the TOF
to form a likelihood value $\mathcal{L}(h)$ for each hadron ($h$)
hypothesis, where $h = p, K$, or $\pi$. Charged tracks are identified
as protons if the proton hypothesis has the highest likelihood
($\mathcal{L}(p) > \mathcal{L}(K)$ and $\mathcal{L}(p) >
\mathcal{L}(\pi)$), or as pions if $\mathcal{L}(\pi) > \mathcal{L}(K)$
is satisfied. The PID efficiencies of protons and pions are both 
approximately 99\% within the momentum range of the signal processes. 
The probability of misidentifying particles as proton is negligible,
and is less than 2\% for misidentification as pions from kaons.

Photon candidates are reconstructed from showers that are not
associated with any charged tracks in the
EMC~\cite{BESIII:2009fln}. The deposited energy of each shower in the
EMC is required to be greater than 25 MeV in the barrel region
($|\cos\theta| < 0.80$), and greater than 50 MeV in the end cap region
($0.86 < |\cos\theta| < 0.92$). The EMC time difference from the event
start time is required to be less than 700 ns, to exclude electronic
noise and showers unrelated to the events. The opening angle between
each shower and $\bar{p}$ must be greater than $20^\circ$, to suppress
the background from annihilation of $\bar{p}$ with the detector
material. The $\piz$ candidates are reconstructed from photon pairs with
invariant mass $M(\gamma\gamma)$ in the range 0.115 $\gevcc$ $<
M(\gamma\gamma) < $ 0.150 $\gevcc$. To improve momentum resolution and
exclude background, a kinematic fit is performed to constrain
$M(\gamma\gamma)$ to the known $\piz$
mass~\cite{ParticleDataGroup:2022pth}, and candidates with fit quality
$\chi^2 < 20$ are retained for further analysis.

The signal candidates of $\lcp \to p\KL$ and $\lcp \to p\KL\piz$ are
required to have only one charged track with opposite charge to the
tagged $\lcm$ satisfying the proton PID criteria. For $\lcp \to
p\KL\piz$ decay, the $\piz$ candidate with the highest energy is
selected. In the reconstruction of $\lcp \to p\KL\pip\pim$, events
must have only three remaining charged tracks with correct charges and
PID. Candidates with additional charged tracks, whose distances of
closest approaches to the IP are within $\pm$20 cm along the beam
direction, are excluded. The presence of the $\KL$ is inferred by the
kinematic variable $\MMsq$, defined as
\begin{equation}
    \MMsq \equiv \left(\ebeam-E_{\mathrm{selected}}\right)^2/c^4 - \left| \vec{p}_{\lcp}-\vec{p}_{\mathrm{selected}} \right|^2/c^2,
\label{eq:MMsq}
\end{equation}
where $\ebeam$ is the beam energy and $E_{\mathrm{selected}}$
($\vec{p}_{\mathrm{selected}}$) is the total measured energy
(momentum) of the selected particles in the DT signal side, boosted
into the c.m. system of $\ee$. To improve the momentum resolution, the
momentum of $\lcp$ is determined by
\begin{equation}
    \vec{p}_{\lcp} \equiv -\hat{p}_{\mathrm{\lcm}}\sqrt{\ebeam^2/c^2 - m^2_{\lcp}c^2},
\label{eq:p_lmdc}
\end{equation}
where $\hat{p}_{\mathrm{\lcm}}$ is the direction of the tagged $\lcm$
and $m_{\lcp}$ is the known $\lcp$ baryon mass taken from the
PDG~\cite{ParticleDataGroup:2022pth}.  For all three decays, the
$\MMsq$ distributions are expected to have a peak around the known
mass squared of $\KL$~\cite{ParticleDataGroup:2022pth}.

Based on studies of inclusive MC samples, the dominant background
events for the signal mode $\lcp \to p\KL\pip\pim$ are from processes
with $\Lambda \to p\pim$ and $\Ks \to \pip\pim$. They are rejected by
vetoing events with $M(p\pim)$ ($M(\pip\pim)$) invariant masses in the
interval of 1.11 $\gevcc < M(p\pim) <$ 1.12 $\gevcc$ (0.48
$\gevcc$$<M(\pip\pim)<$ 0.52 $\gevcc$).  The combinatorial backgrounds
are suppressed by requiring the recoil mass of the proton
$M_{\mathrm{recoil}}(p) \equiv \sqrt{E_\mathrm{beam}^2 -
  |\vec{p}_{\lcp} - \vec{p}_p|^2} > 1.0$ $\gevcc$, which removes only
about 3\% of the signal. Here $\vec{p}_p$ is the momentum of the
proton.  For the $\lcp \to p\KL\piz$ signal mode, background events of
$\lcp \to p\Ks(\to\piz\piz)$ and $p\KL$ are excluded by requiring
$M_{\mathrm{recoil}}(p) > 0.65 \gevcc$, which removes less than 1\% of
signal.  Events within the range 1.17 $\gevcc < M(p\piz) < 1.20$
$\gevcc$ are discarded to suppress the background of the $\Sigma^+ \to
p\piz$ decay.

To improve the momentum resolution, a six constraint (6C) kinematic
fit is performed requiring total four-momentum conservation with
respect to that of the initial $\ee$ collision and constraining both
masses of the tagged $\lcm$ and the signal $\lcp$ to
$m_{\lcp}$. The $\KL$ is treated as a missing particle, and its
four-momentum and mass are free in the kinematic fit. The $\chi^2$
of the kinematic fit for each signal mode is required to be less than
the optimized value that maximizes the figure of merit $S/\sqrt{S+B}$,
where $S$ and $B$ are the numbers of signal and background events from 
the inclusive MC samples, scaled to the data luminosity. Here, the BF 
of the signal modes are assumed to be the same as the measured ones of $\lcp \to \Ks X$.
The optimized requirements are $\chi^2 < 60$ for $\lcp \to p\KL$, $\chi^2 < 25$ for
$\lcp \to p\KL\pip\pim$, and $\chi^2 < 20$ for $\lcp \to p\KL\piz$.
The resulting $\MMsq$ distributions of the DT events are shown in
Figure~\ref{fig:mm2}, which combine all data samples at the seven
c.m. energies. Signal events are indicated by the significant peaks
around the $\KL$ mass squared. 

\begin{figure}[!hpt]
    \centering
    \includegraphics[width=0.48\textwidth]{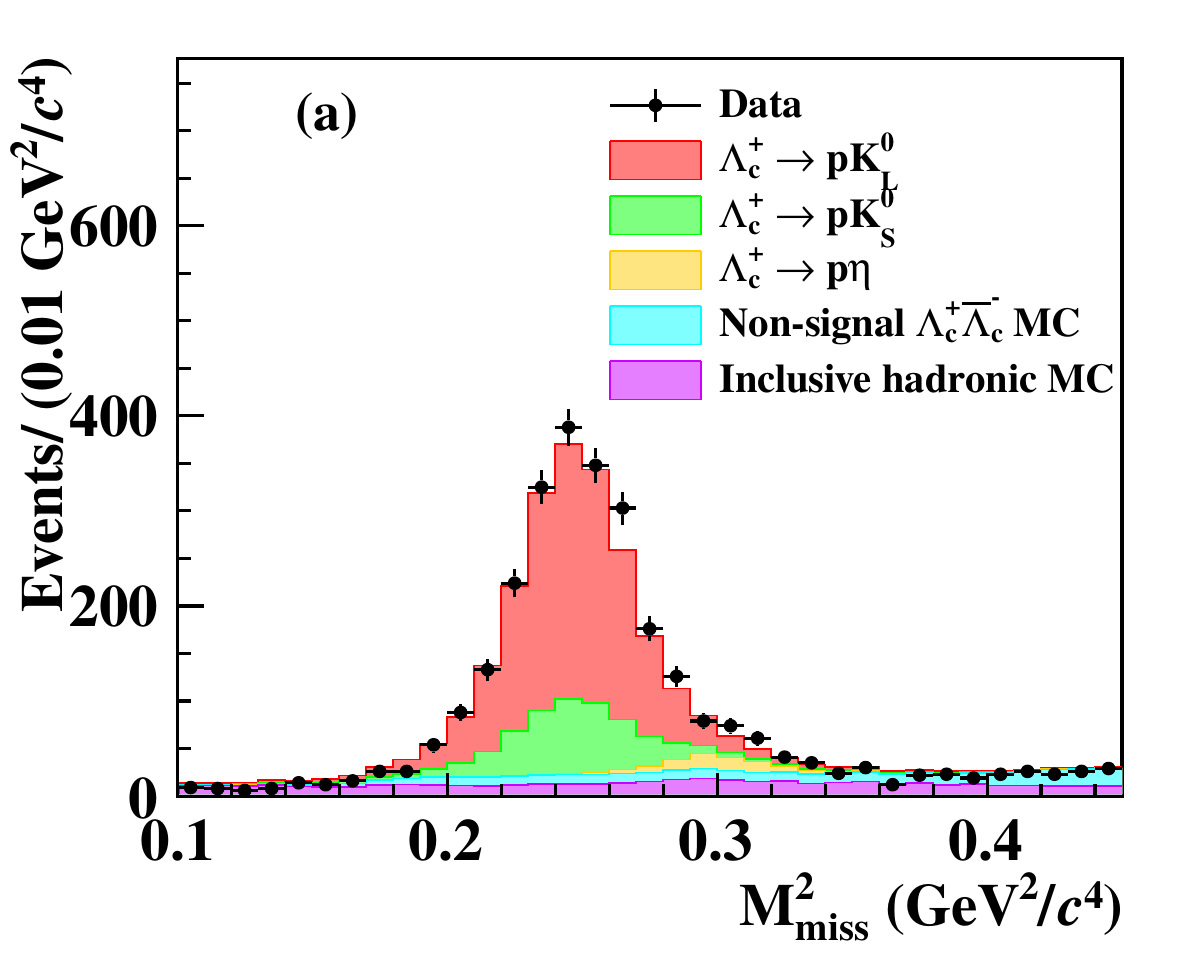}
    \includegraphics[width=0.48\textwidth]{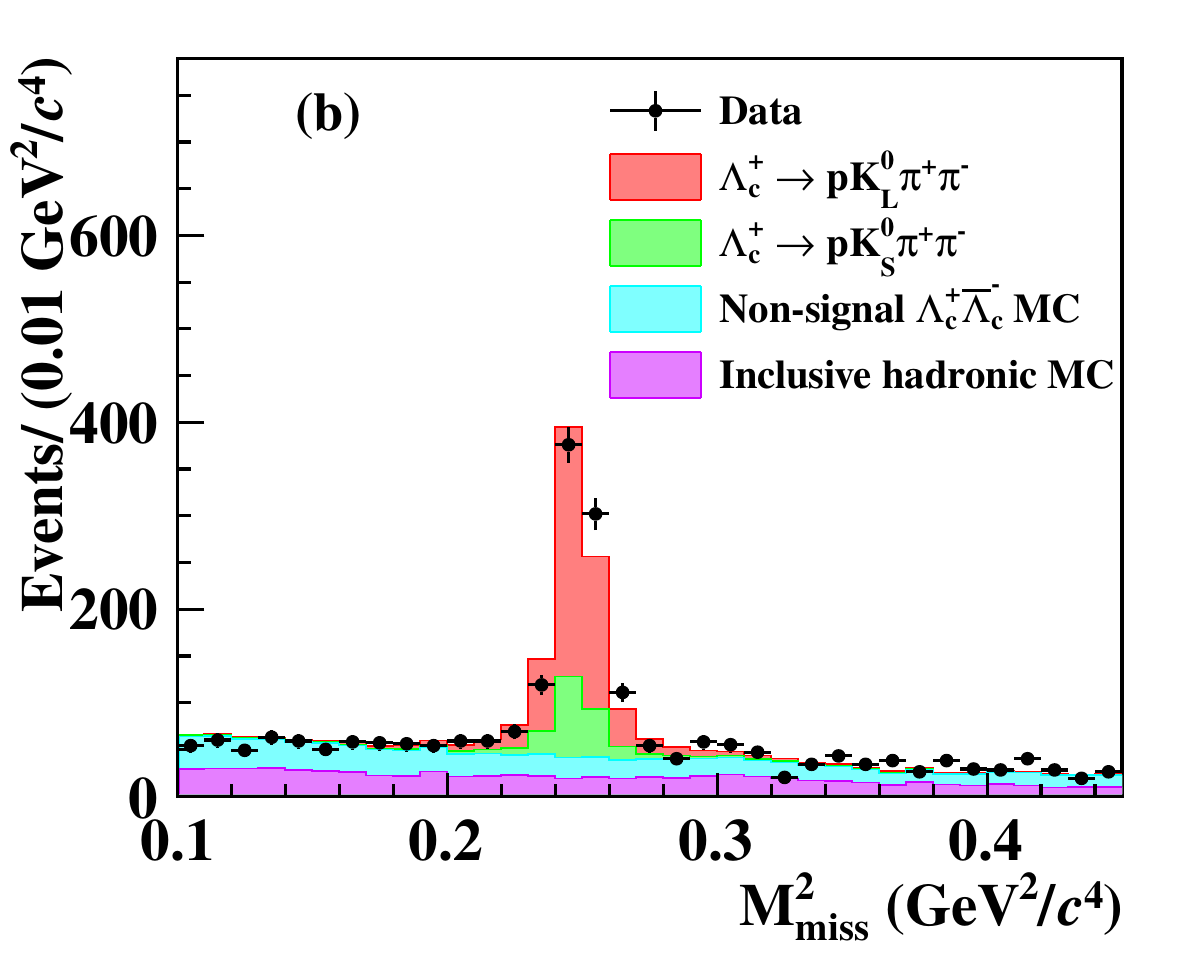} \\
    \includegraphics[width=0.48\textwidth]{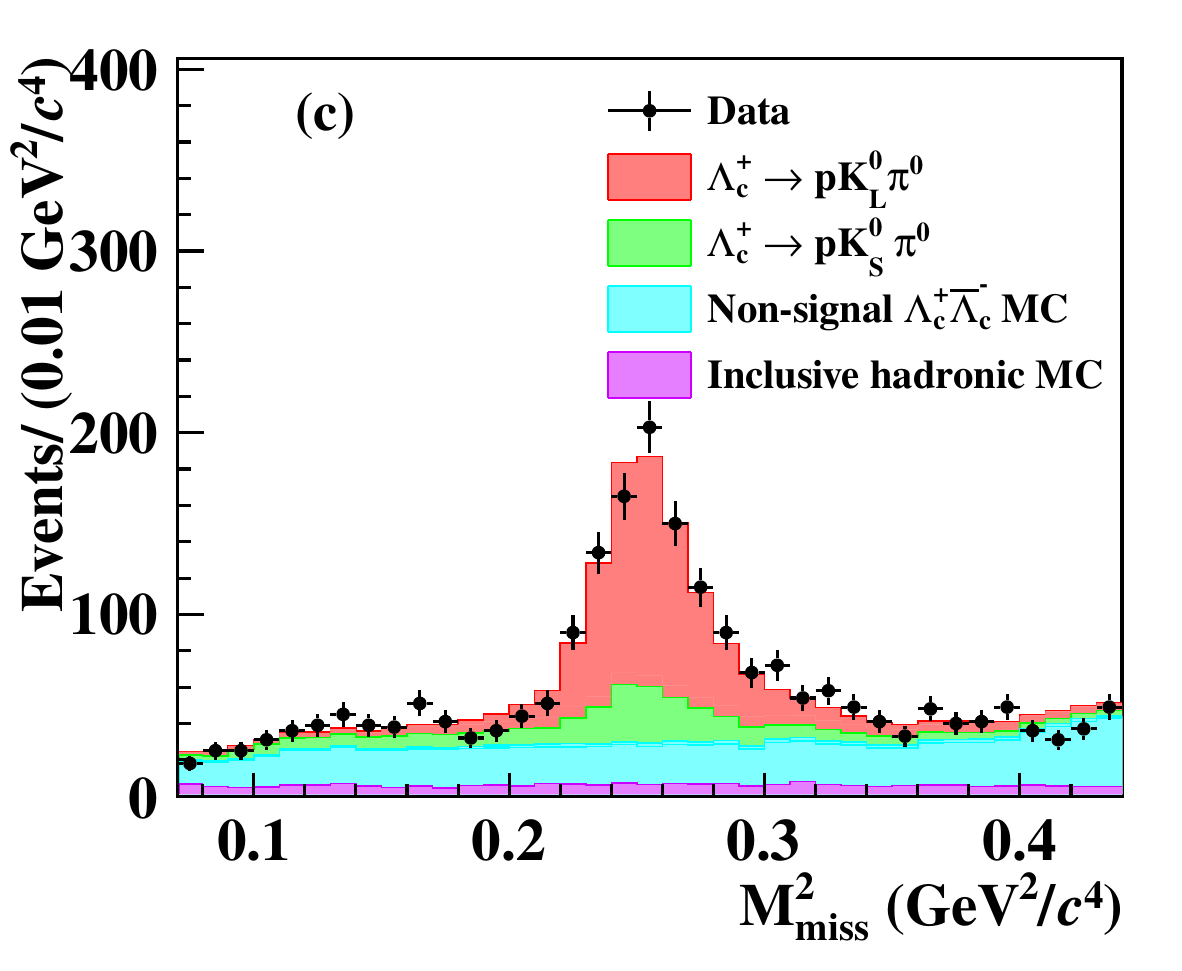}
    \caption{The $\MMsq$ distributions of the selected DT events for
      (a) $\lcp \to p\KL$, (b) $\lcp \to p\KL\pip\pim$, and (c) $\lcp
      \to p\KL\piz$ decays. The points with error bars are data
      combined from seven c.m. energies, the red histograms indicate
      the signal processes, and the green and orange histograms are the
      peaking backgrounds. The cyan and violet histograms represent
      non-signal $\lcp\lcm$ and inclusive hadronic background
      processes, respectively.}
    \label{fig:mm2}
\end{figure}

There are peaking backgrounds remaining from $\lcp \to
p\Ks(\to\piz\piz)$ and $\lcp \to p\eta(\to\gamma\gamma$ or $3\piz)$,
$\lcp \to p\Ks(\to\piz\piz)\pip\pim$, and $\lcp \to
p\Ks(\to\piz\piz)\piz$ in the corresponding signal modes.  The peaking
background events from $\lcp \to \Ks X$ decays $N^{\mathrm{Bkg}}_{\Ks
  X}$ are determined by
\begin{eqnarray}
    N^{\mathrm{Bkg}}_{\Ks X}=N^{\mathrm{Data}}_{\mathrm{DT} \,\Ks X}\cdot w_{\Ks X},\ w_{\Ks X} = \frac{\sum_{i}s_i \cdot \frac{\NST_i}{\effST_i}\cdot N^{\mathrm{MC},i}_{\mathrm{DT}\,\KL X}}{\sum_i\frac{\NST_i}{\effST_i}\cdot N^{\mathrm{MC},i}_{\mathrm{DT} \,\Ks X}},
    \label{eq:nbkg_pKsX}
\end{eqnarray}
where $i$ represents the tag mode, and $N^{\mathrm{Data}}_{\mathrm{DT}
  \,\Ks X}$ denotes the data yields passing the DT selection criteria
of $\lcp \to \Ks X$. Here, the DT selection criteria of $\lcp \to \Ks
X$ require a fully reconstructed $\Ks$ from $\pip\pim$ combinations,
as described in Ref.~\cite{BESIII:2015bjk}.
$N^{\mathrm{Data}}_{\mathrm{DT} \,\Ks X}$ is corrected by the factor
$w_{\Ks X}$, which is derived from the exclusive MC simulation samples
of $\lcp \to \Ks X$. $N_{\mathrm{DT}\,\KL X}^{\mathrm{MC},i}$ and
$N_{\mathrm{DT}\,\Ks X}^{\mathrm{MC},i}$ are the numbers of the $\Ks
X$ MC events that satisfy the DT selection criteria of $\lcp \to
\KL X$ and $\lcp \to \Ks X$, respectively. $\NST_i$ and $\effST_i$ are
the ST yields and ST efficiencies from Ref.~\cite{BESIII:2022xne}. A
scale factor $s_i$ is specified for each tag mode, and $s_i$ is set to
2 if both the tag and signal modes are $\Ks X$. Otherwise, it is set
to 1. For peaking background events from $\lcp \to p\eta$, the
contribution is evaluated based on the corresponding exclusive MC
samples using
\begin{eqnarray}
    N^{\mathrm{Bkg}}_{p\eta}=\BR(\lcp\to p \eta)\cdot w_{p\eta},\ w_{p\eta} =\sum_i\left(  \frac{\NST_i}{\effST_i}\cdot \frac{N'^{\mathrm{MC},i}_{p\KL}}{N'^{\mathrm{MC},i}_{\mathrm{tot}}} \right),
    \label{eq:nbkg_pEta}
\end{eqnarray}
with $\BR(\lcp\to
p\eta)=(1.41\pm0.11)\times10^{-3}$~\cite{ParticleDataGroup:2022pth}. $N'^{\mathrm{MC},i}_{p\KL}$
is the number of surviving DT events for the $i$-th tag mode, that
satisfy the DT selection criteria of $\lcp\to p \KL$, and
$N'^{\mathrm{MC},i}_{\mathrm{tot}}$ is the total number of MC events
generated for the $i$-th tag mode. Table~\ref{tab:peaking_yields}
summarizes the contributions arising from each peaking background
process.

\begin{table}[!htbp]
	\begin{center}
\caption{Estimated yields of peaking backgrounds at each
  c.m. energy. The uncertainties are statistical only.}
\label{tab:peaking_yields}
\begin{tabular}{l|c|c|c|c}
\hline \hline
$\sqrt{s}$ (GeV) & $\lcp \to p\Ks$ & $\lcp \to p\eta$ & $\lcp \to p\Ks\pip\pim$ & $\lcp \to  p\Ks\piz$ \\\hline
4.600 & 59 $\pm$ 6   & 13 $\pm$ 1   & 25 $\pm$ 5  & 44 $\pm$ 6 \\
4.612 & 14 $\pm$ 3   & 2.3$\pm$ 0.2 & 4  $\pm$ 1  & 8 $\pm$ 3  \\
4.628 & 70 $\pm$ 7   & 11 $\pm$ 1   & 31 $\pm$ 4  & 38 $\pm$ 6 \\
4.641 & 68 $\pm$ 7   & 12 $\pm$ 1   & 30 $\pm$ 5  & 38 $\pm$ 7 \\
4.661 & 60 $\pm$ 6   & 12 $\pm$ 1   & 39 $\pm$ 5  & 51 $\pm$ 8 \\
4.682 & 198 $\pm$ 12 & 35 $\pm$ 3   & 97 $\pm$ 9  & 150 $\pm$ 13 \\
4.699 & 54 $\pm$ 6   & 10 $\pm$ 1   & 24 $\pm$ 5  & 37 $\pm$ 6 \\
\hline \hline 
 \end{tabular}	
\end{center}
\end{table}

A simultaneous unbinned maximum-likelihood fit is performed on the
$\MMsq$ distributions of the seven c.m. energies. The signal and
peaking backgrounds are modeled by individual MC-simulated shapes
convolved with Gaussian functions to account for differences
between the data and MC simulations. The Gaussian means and widths are
free parameters in the fit.  The yields of the peaking background
events are free with their mean
and standard deviation values set to the
results listed in Table~\ref{tab:peaking_yields}. 
For the signal mode $\lcp \to p\KL\piz$, a truth-matching method is
employed to obtain the pure signal shape by comparing the two photons
from the $\piz$ with their corresponding MC truth information. The
opening angle $\theta_{\mathrm{truth}}$ between the truth and the
reconstructed photons is required to be less than $10^{\circ}$.  The
combinatorial background shape is taken from the inclusive MC samples,
including non-signal $\lcp\lcm$ and continuum hadron production
events.

The BFs of the decays $\lcp \to p\KL$, $\lcp \to p\KL\pip\pim$, and
$\lcp \to p\KL\piz$ are shared variables for the seven
c.m. energies in the simultaneous fit, determined by
\begin{eqnarray}
    \BR_\mathrm{sig}=\frac{\NDT}{\NST \cdot \varepsilon_{\mathrm{avg}} \cdot \BR_{\mathrm{int}}} ,
    \label{eq:BFsig}
\end{eqnarray}
where $\varepsilon_{\mathrm{avg}} = \left(\sum_i
\NST_i\cdot\effDT_i/\effST_i\right)/\NST$ is the average detection
efficiency for detecting signal modes in ST events and $i$ represents
the $i$-th ST tag mode. Table~\ref{tab:avg_dt_eff} lists the ST events
and the average detection efficiencies for each c.m. energy. $\NDT$
and $\effDT_i$ are the DT yields and corresponding efficiencies,
respectively. $\BR_{\mathrm{int}}$ is the intermediate BF of $\piz$,
$\BR(\piz \to \gamma\gamma) =
(98.823\pm0.034)\%$~\cite{ParticleDataGroup:2022pth} for $\lcp \to
p\KL\piz$ decay.
Figure~\ref{fig:DT_fit} shows the results of fits to the $\MMsq$
distributions, combining all data samples. From these fits, the BFs
are  $\BR(\lcp \to p\KL) = (1.67\pm0.06)\%$, $\BR(\lcp
\to p\KL\pip\pim) = (1.69\pm0.10)\%$, and $\BR(\lcp \to p\KL\piz) =
(2.02\pm0.13)\%$, where the uncertainties are statistical only. The
total DT signal yields from all c.m. energies are
$N^{\mathrm{DT}}_{p\KL}=1627\pm56$,
$N^{\mathrm{DT}}_{p\KL\pip\pim}=648\pm39$, and
$N^{\mathrm{DT}}_{p\KL\piz}=652\pm41$, for $\lcp \to p\KL$, $\lcp \to
p\KL\pip\pim$, and $\lcp \to p\KL\piz$, respectively.
\begin{table}[H]
\begin{center}
\caption{ST events~($\NST$), average detection efficiencies~($\varepsilon_{\mathrm{avg}}$) and DT
  signal yields~($\NDT$) for $\lcp \to p\KL$, $\lcp \to p\KL\pip\pim$, and
  $\lcp \to p\KL\piz$ decays at each c.m. energy. The errors are
  statistical only.}
\label{tab:avg_dt_eff}
\resizebox{1.0\textwidth}{!}{
\begin{tabular}{l|c|c|c|c|c|c|c}
\hline \hline
 & $4.600~\gev$ & $4.612~\gev$  & $4.628~\gev$  & $4.641~\gev$ & $4.661~\gev$ & $4.682~\gev$ & $4.698~\gev$ \\ \hline
$N^\mathrm{ST}$ & 17391 $\pm$ 171 & 3114 $\pm$ 75 & 14558 $\pm$ 135 & 15545 $\pm$ 165 & 15235 $\pm$ 164 & 44704 $\pm$ 284 & 12971 $\pm$ 158\\\hline
modes & \multicolumn{7}{c}{$\varepsilon_{\mathrm{avg}} (\%)$} \\\hline
$\lcp \to p\KL$ & 76.32 $\pm$ 0.08 & 76.13 $\pm$ 0.08 & 78.12 $\pm$ 0.08 & 78.83 $\pm$ 0.08 & 78.78 $\pm$ 0.08 & 79.66 $\pm$ 0.09 & 79.89 $\pm$ 0.09\\
$\lcp \to p\KL\pip\pim$ & 28.61 $\pm$ 0.16 & 28.86 $\pm$ 0.17 & 29.87 $\pm$ 0.18 & 30.96 $\pm$ 0.17 & 31.12 $\pm$ 0.17 & 31.93 $\pm$ 0.17 & 32.83 $\pm$ 0.18\\
$\lcp \to p\KL\piz$ & 23.83 $\pm$ 0.15 & 24.28 $\pm$ 0.15 & 25.79 $\pm$ 0.16 & 26.40 $\pm$ 0.16 & 26.68 $\pm$ 0.16 & 27.43 $\pm$ 0.17 & 27.27 $\pm$ 0.17\\\hline
modes & \multicolumn{7}{c}{$N^{\mathrm{DT}}$} \\\hline
$\lcp \to p\KL$ & 222 $\pm$ 8 & 40 $\pm$ 1 & 190 $\pm$ 7 & 205 $\pm$ 7 & 201 $\pm$ 7 & 596 $\pm$ 21 & 173 $\pm$ 6 \\\hline
$\lcp \to p\KL\pip\pim$ & 84 $\pm$ 5 & 15 $\pm$ 1 & 74 $\pm$ 4 & 81 $\pm$ 5 & 80 $\pm$ 5 & 242 $\pm$ 14 & 72 $\pm$ 4 \\\hline
$\lcp \to p\KL\piz$ & 83 $\pm$ 5 & 15 $\pm$ 1 & 75 $\pm$ 5 & 82 $\pm$ 5 & 81 $\pm$ 5 & 245 $\pm$ 16 & 71 $\pm$ 4 \\
\hline \hline
\end{tabular}	
} 
\end{center}
\end{table}

\begin{figure}[H]\centering
\includegraphics[width=0.48\textwidth]{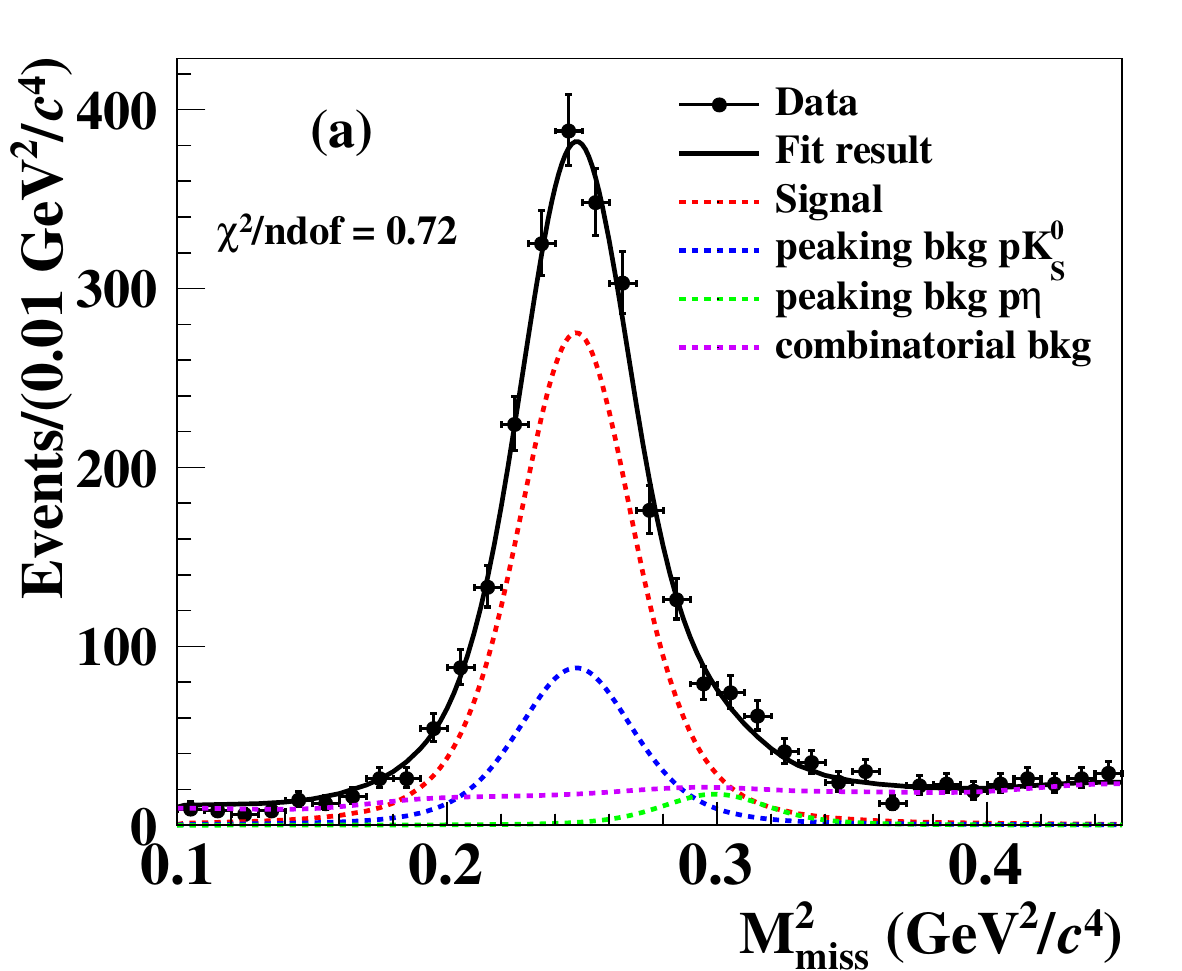}
\includegraphics[width=0.48\textwidth]{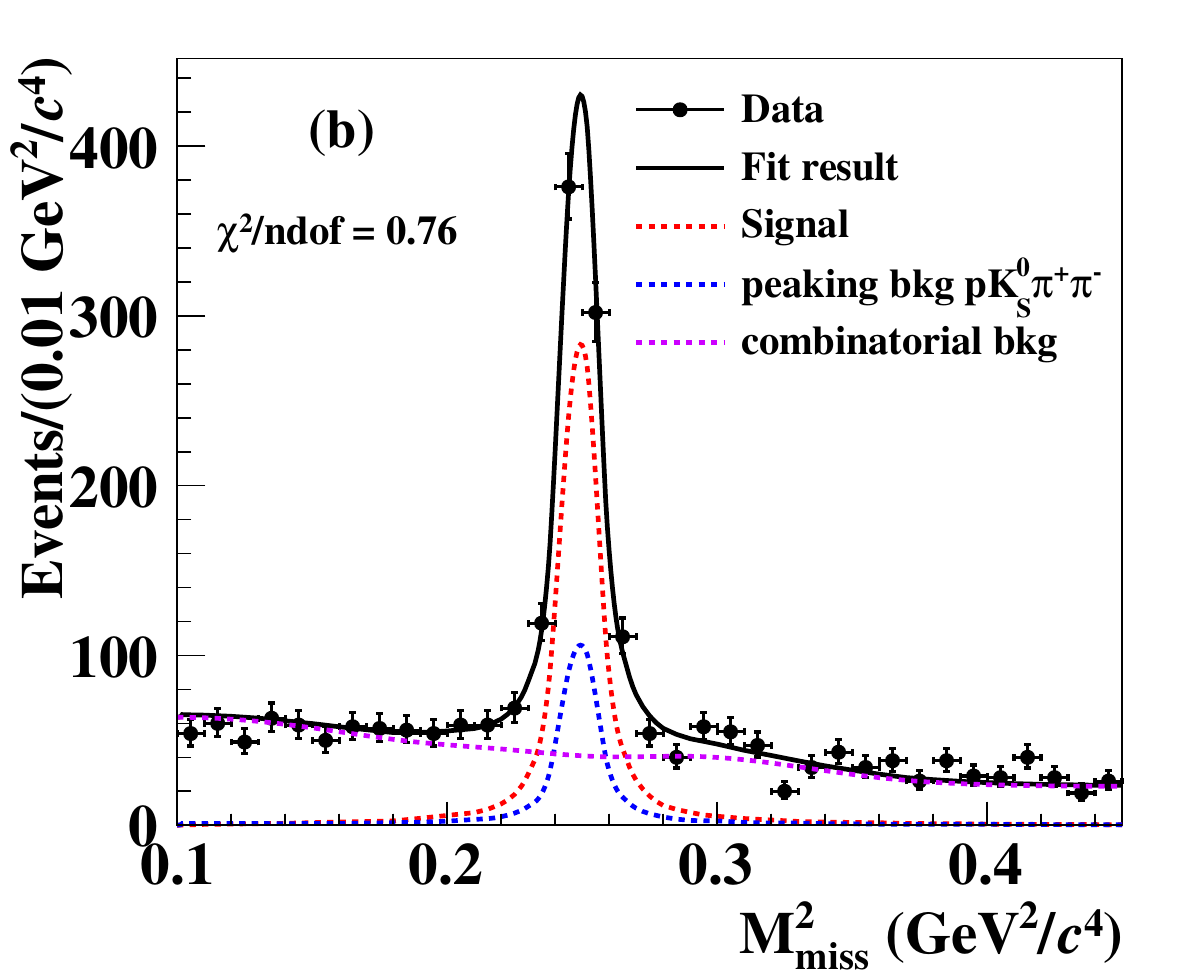} \\
\includegraphics[width=0.48\textwidth]{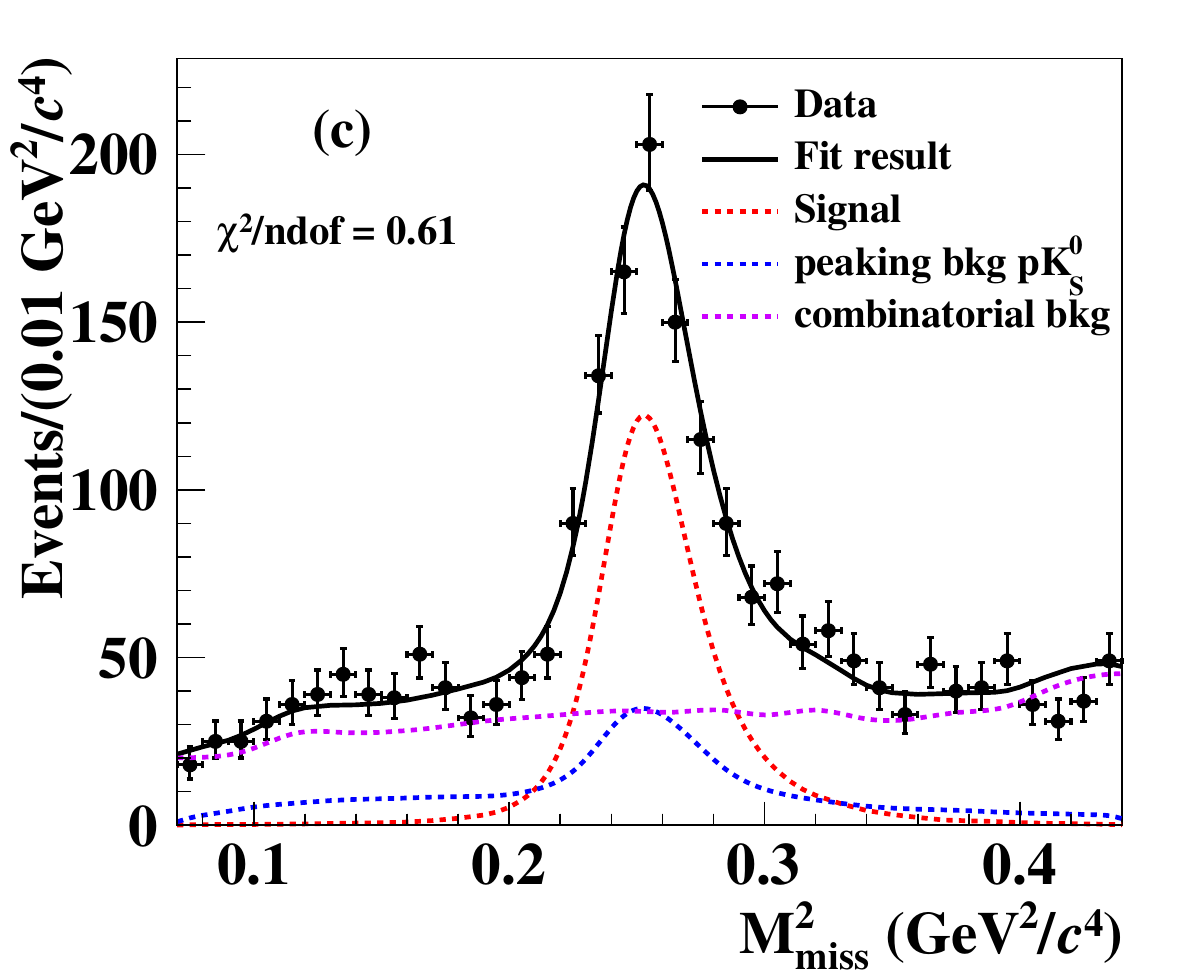}
\caption{Combined fit results of the $\MMsq$ distributions from all
  data samples for (a) $\lcp \to p\KL$, (b) $\lcp \to p\KL\pip\pim$
  and (c) $\lcp \to p\KL\piz$ decays. The black dots with error bars
  are data, while the black solid curves are the fit results. The red
  dashed curves are the signal shapes, and the blue and green dashed curves
  represent the peaking backgrounds $\lcp \to p\Ks$ and $\lcp \to
  p\eta$, respectively. The violet dashed curves are the combinatorial
  background shapes.}
\label{fig:DT_fit}
\end{figure}

\section{Systematic uncertainties}
\label{sec:systematic}
\hspace{1.5em} In the DT method, most of the systematic uncertainties
associated with the ST selections cancel. The major sources of
systematic uncertainties in the BFs measurements are described below
and are reported relative to the measured BFs.

\begin{itemize}
\item \textbf{Tracking and PID efficiencies}. The tracking and PID
  efficiencies of the charged protons and pions are studied using a
  control sample of $\jpsi \to
  p\bar{p}\pip\pim$~\cite{BESIII:2022bkj}. The MC simulation samples
  are weighted by the efficiency ratio between data and MC as function
  of charged particle momentum and $\cos{\theta}$.  The systematic
  uncertainties of tracking and PID are 0.5\% and 0.1\% for
  $\lcp \to p\KL$, 1.6\% and 0.8\% for $\lcp \to p\KL\pip\pim$, and
  0.7\% and 0.4\% for $\lcp \to p\KL\piz$, respectively.
\item \textbf{No extra charged track requirement}. The number of good
  charged tracks is required to be exactly one (three) for $p\KL$ and
  $p\KL\piz$ ($p\KL\pip\pim$) DT candidates in the recoil system of
  the tagged $\lcm$. The difference between data and MC simulation
  from this selection is studied using a control sample of $\lcp \to
  pK^+\pim$. The systematic uncertainty is 1.9\%.
\item \textbf{MC statistics}. The exclusive MC simulation samples are
  used to obtain the ST and DT detection efficiencies and to estimate
  the peaking background events. The systematic uncertainties
  associated with the limited MC sample sizes are estimated to be
  0.1\%, 0.5\%, and 0.5\% for $\lcp \to p\KL$, $\lcp \to
  p\KL\pip\pim$, and $\lcp \to p\KL\piz$, respectively.

\item \textbf{ST yield}. The systematic uncertainty arising from the
  total ST yield is assigned to be 0.2\%~\cite{BESIII:2022xne}.

\item \textbf{Kinematic fit}. The model of the MC simulation is much
  simpler than the real detector performance, resulting in a
  difference between the data and MC simulation in the track
  parameters of the charged tracks~\cite{BESIII:2012mpj}. 
  The correction parameters are obtained through a data-driven method
  using control samples of $\ee \to p\bar{p}\pip\pim$ and $\ee \to K^+K^-\pip\pim$.
  The helix parameters of the charged tracks are corrected, and the BFs are
  re-evaluated with the updated MC simulation samples. The differences
  from the measured BFs are taken as the systematic uncertainties
  associated with the kinematic fit, which are 0.5\%, 1.0\%, and 0.5\%
  for $\lcp \to p\KL$, $\lcp \to p\KL\pip\pim$ and $\lcp \to
  p\KL\piz$, respectively.

\item \textbf{\boldmath Angle($\gamma$,$\bar{p}$) requirement}. To estimate
  the systematic uncertainty of the Angle($\gamma, \bar{p} $) requirement, the
  difference between the data and MC simulation samples of this
  requirement is investigated from a control sample of $\psi(3686) \to
  \pi^+\pi^-\jpsi, \jpsi \to p \bar{p} \pi^0$. The systematic
  uncertainty is 0.2\% for $\lcp \to p\KL\piz$.

\item \textbf{\boldmath $\piz$ reconstruction}. The systematic uncertainty due
  to the $\piz$ reconstruction is determined using the control sample
  of $\jpsi \to p \bar{p} \piz$~\cite{BESIII:2022ahw}. The MC
  simulation samples are corrected depending on the $\piz$
  momentum. The systematic uncertainty is determined to be 0.5\%.

\item \textbf{Truth-match method}. The systematic uncertainty from the
  truth-match method is determined comparing the measured BFs with and
  without the truth-match requirements. The resulting systematic
  uncertainty is taken as 0.2\%.

\item \textbf{Signal model}. For $\lcp \to p\KL$, the systematic
  uncertainty from the signal model is determined varying the decay
  asymmetry parameters within $\pm 1\sigma$. The deviation from the
  measured BF is found to be negligible. For
  $\lcp \to p\KL\pip\pim$ and $\lcp \to p\KL\piz$, the signal models
  in the nominal analysis is tunned based on the data.
  The possible intermediate resonances are considered in the amplitude analysis, 
  composed of $\Sigma^*$, $\Delta^*$, $N^*$, $\bar K^*$ and $\rho$. 
  The nominal amplitude models are then replaced by alternative ones
  with equivalent descriptions of the data. 
  The alternative model of $\lcp \to p\KL\piz$ is selected by including other additional
  insignificant resonance. In the case of $\lcp \to p\KL\pip\pim$, the amplitude fit 
  is not stable due to the limited statistics of data. The same components of resonances 
  are employed as in the nominal case, with a different fit solution that yields a negative log-likelihood value agreeing closely with the nominal one. 
  The systematic uncertainties are determined to be 1.1\%
  and 0.9\% for $\lcp \to p\KL\pip\pim$ and $\lcp \to p\KL\piz$,
  respectively.

\item \textbf{Background shape}. To investigate the systematic
  uncertainty from the background shape, the nominal background shape
  is replaced with a second-order Chebychev polynomial function in the
  simultaneous fit. The systematic uncertainties are
  0.9\%, 0.6\%, and 0.4\% for $\lcp \to p\KL$, $\lcp \to p\KL\pip\pim$
  and $\lcp \to p\KL\piz$, respectively.

\item \textbf{Fit bias}. The systematic uncertainty from the
  simultaneous fit is studied with 5000 sets of toy MC samples, which are
  simulated with all parameters from the fit model fixed. The BFs
  obtained from the toy samples are fitted with a Gaussian function. The
  deviations between the Gaussian mean value and nominal BFs are
  assigned as systematic uncertainties. For the decay $\lcp
  \to p\KL\piz$, the fit bias is found to be 0.3\%, while for the
  other two signal modes it is negligible.
\end{itemize}

Other sources of systematic uncertainties, such as the BF of
$\piz\to\gamma\gamma$, are neglected due to their negligible
effects. Assuming that all sources of systematic uncertainties in the
BFs measurements are uncorrelated, the quadratic sums of the different
sources are considered as the total systematic uncertainties, which are
2.2\%, 3.1\%, and 2.5\% for $\lcp \to p\KL$, $\lcp \to p\KL\pip\pim$,
and $\lcp \to p\KL\piz$, respectively. Table~\ref{tab:sys_sources}
lists all the systematic uncertainties discussed above.
\begin{table}[htbp]
\begin{center}
\caption{Relative systematic uncertainties in the BF measurements.}
\footnotesize
\begin{tabular}{l | c | c | c }
	\hline \hline
	Source & $\lcp \to p\KL \ (\%)$ &  $\lcp \to p\KL\pip\pim \ (\%)$ & $\lcp \to p\KL\piz \ (\%)$\\
	\hline
	Tracking                           & 0.5  &   1.6  & 0.7 \\
	PID                                & 0.1  &   0.8  & 0.4 \\ 
	No extra charged track             & 1.9  &   1.9  & 1.9 \\
	MC statistics                      & 0.2  &   0.5  & 0.5 \\
	ST  yields                         & 0.2  &   0.2  & 0.2 \\
	Kinematic fit                      & 0.5  &   1.0  & 0.5 \\
	Angle($\gamma$,$\bar{p}$) requirement & -    &    -   & 0.2 \\
	$\piz$ reconstruction              & -    &    -   & 0.5 \\ 
    Truth-match method                 & -    &    -   & 0.2 \\ 
	Signal model                       & -    &   1.1  & 0.9 \\
	Background shape                   & 0.9  &   0.6  & 0.4 \\ 
    Fit bias                           & -    &   -    & 0.3 \\ 
 \hline 
	Total                              & 2.2  &   3.1  & 2.5 \\
	\hline\hline
\end{tabular}
\label{tab:sys_sources}
\end{center}
\end{table}

\section{Summary}
\label{sec:summary}
\hspace{1.5em} In summary, we report the BFs of $\lcp \to p\KL$, $\lcp
\to p\KL\pip\pim$ and $\lcp \to p\KL\piz$ for the first time, by
analyzing $\ee$ annihilation data samples corresponding to an integrated
luminosity of 4.5 $\ifb$ collected at c.m. energies between 4.600 and
4.699 $\gev$. The measured BFs of these decays are $\BR(\lcp \to p\KL)
= (1.67 \pm 0.06 \pm 0.04)\%$, $\BR(\lcp \to p\KL\pip\pim) = (1.69 \pm
0.10 \pm 0. 05)\%$, and $\BR(\lcp \to p\KL\piz) = (2.02 \pm 0.13 \pm
0.05)\%$. Combining the BFs measurements in this work with the
values of $\BR(\lcp \to \Ks X)$~\cite{ParticleDataGroup:2022pth}, the
$\Ks$-$\KL$ asymmetries are determined, as summarized in
Table~\ref{tab:results}. The uncertainties are derived through the
standard error propagation procedure, assuming that the uncertainties
of the estimated $\BR(\lcp \to \KL X)$ and the quoted $\BR(\lcp \to
\Ks X)$ are uncorrelated.  Taking into account the uncertainties, no
obvious asymmetry is observed in any of the three decays. The
$\Ks$-$\KL$ asymmetry of $\lcp \to pK_{S,L}^0$ $R(\lcp, pK_{S,L}^0) =
-0.025 \pm 0.031$ is compatible with the prediction of
($-$0.010, 0.087) based on SU(3) flavor
symmetry~\cite{Wang:2017gxe}. Our
measurements of the $\Ks$-$\KL$ asymmetries in charmed baryon decays
offer the possibility to access the DCS processes involving neutral
kaons and provide further constraints on their amplitudes.

\begin{table}[H]
    \begin{center}
    \caption{The BFs $\BR(\lcp \to \KL X)$, the known BFs $\BR(\lcp \to \Ks X)$, and $\Ks$-$\KL$ asymmetries.}
    \label{tab:results}
    \resizebox{1.0\textwidth}{!}{
    \begin{tabular}{l|c|c|c}
    \hline \hline
    Mode      & $\BR(\lcp \to \KL X)$ (\%)  & $\BR(\lcp \to \Ks X)$ (\%)~\cite{ParticleDataGroup:2022pth} & $R(\lcp$, $K_{L,S}^{0}X)$ \\ \hline
    $\lcp \to pK_{L,S}^{0}$           & 1.67 $\pm$ 0.06 $\pm$ 0.04 & 1.59 $\pm$ 0.07 & -0.025 $\pm$ 0.031 \\
    $\lcp \to pK_{L,S}^{0}\pip\pim$   & 1.69 $\pm$ 0.10 $\pm$ 0.05 & 1.60 $\pm$ 0.11 & -0.027 $\pm$ 0.048 \\
    $\lcp \to pK_{L,S}^{0}\piz$       & 2.02 $\pm$ 0.13 $\pm$ 0.05 & 1.96 $\pm$ 0.12 & -0.015 $\pm$ 0.046 \\
    \hline \hline
    \end{tabular}}
    \end{center}
\end{table}

\acknowledgments
\hspace{1.5em}
The BESIII Collaboration thanks the staff of BEPCII and the IHEP computing center for their strong support. This work is supported in part by National Key R\&D Program of China under Contracts Nos. 2020YFA0406400, 2020YFA0406300, 2023YFA1606000; National Natural Science Foundation of China (NSFC) under Contracts Nos. 11635010, 11735014, 11935015, 11935016, 11935018, 12025502, 12035009, 12035013, 12061131003, 12192260, 12192261, 12192262, 12192263, 12192264, 12192265, 12221005, 12225509, 12235017, 12361141819; the Chinese Academy of Sciences (CAS) Large-Scale Scientific Facility Program; the CAS Center for Excellence in Particle Physics (CCEPP); Joint Large-Scale Scientific Facility Funds of the NSFC and CAS under Contract No. U1832207; 100 Talents Program of CAS; The Institute of Nuclear and Particle Physics (INPAC) and Shanghai Key Laboratory for Particle Physics and Cosmology; German Research Foundation DFG under Contracts Nos. 455635585, FOR5327, GRK 2149; Istituto Nazionale di Fisica Nucleare, Italy; Ministry of Development of Turkey under Contract No. DPT2006K-120470; National Research Foundation of Korea under Contract No. NRF-2022R1A2C1092335; National Science and Technology fund of Mongolia; National Science Research and Innovation Fund (NSRF) via the Program Management Unit for Human Resources \& Institutional Development, Research and Innovation of Thailand under Contract No. B16F640076; Polish National Science Centre under Contract No. 2019/35/O/ST2/02907; The Swedish Research Council; U. S. Department of Energy under Contract No. DE-FG02-05ER41374

\bibliographystyle{JHEP}
\bibliography{main.bib}

\providecommand{\href}[2]{#2}\begingroup\raggedright\begin{thebibliography}{10}

\bibitem{Cheng:2021qpd}
H.-Y.~Cheng, \emph{{Charmed baryon physics circa 2021}}, \href{https://doi.org/10.1016/j.cjph.2022.06.021}{\emph{Chin. J. Phys.} {\bfseries 78} (2022) 324} [\href{https://arxiv.org/abs/arXiv:2109.01216}{{\ttfamily arXiv:2109.01216}}].

\bibitem{Korner:1992wi}
J.G.~Korner and M.~Kramer, \emph{{Exclusive nonleptonic charm baryon decays}}, \href{https://doi.org/10.1007/BF01561305}{\emph{Z. Phys. C} {\bfseries 55} (1992) 659}.

\bibitem{Ivanov:1997ra}
M.A.~Ivanov, J.G.~Korner, V.E.~Lyubovitskij and A.G.~Rusetsky, \emph{{Exclusive nonleptonic decays of bottom and charm baryons in a relativistic three quark model: Evaluation of nonfactorizing diagrams}}, \href{https://doi.org/10.1103/PhysRevD.57.5632}{\emph{Phys. Rev. D} {\bfseries 57} (1998) 5632} [\href{https://arxiv.org/abs/hep-ph/9709372}{{\ttfamily hep-ph/9709372}}].

\bibitem{Cheng:1991sn}
H.-Y.~Cheng and B.~Tseng, \emph{{Nonleptonic weak decays of charmed baryons}}, \href{https://doi.org/10.1103/PhysRevD.46.1042}{\emph{Phys. Rev. D} {\bfseries 46} (1992) 1042}.

\bibitem{Xu:1992vc}
Q.P.~Xu and A.N.~Kamal, \emph{{Cabibbo favored nonleptonic decays of charmed baryons}}, \href{https://doi.org/10.1103/PhysRevD.46.270}{\emph{Phys. Rev. D} {\bfseries 46} (1992) 270}.

\bibitem{Cheng:1993gf}
H.-Y.~Cheng and B.~Tseng, \emph{{Cabibbo allowed nonleptonic weak decays of charmed baryons}}, \href{https://doi.org/10.1103/PhysRevD.48.4188}{\emph{Phys. Rev. D} {\bfseries 48} (1993) 4188} [\href{https://arxiv.org/abs/hep-ph/9304286}{{\ttfamily hep-ph/9304286}}].

\bibitem{Xu:1992sw}
Q.P.~Xu and A.N.~Kamal, \emph{{The Nonleptonic charmed baryon decays: $B_c \to B(\frac{3}{2}^+,\,\mathrm{decuplet})+P(0^-)$ or $V(1^-)$}}, \href{https://doi.org/10.1103/PhysRevD.46.3836}{\emph{Phys. Rev. D} {\bfseries 46} (1992) 3836}.

\bibitem{Zenczykowski:1993jm}
P.~Zenczykowski, \emph{{Nonleptonic charmed baryon decays: Symmetry properties of parity violating amplitudes}}, \href{https://doi.org/10.1103/PhysRevD.50.5787}{\emph{Phys. Rev. D} {\bfseries 50} (1994) 5787}.

\bibitem{Sharma:1998rd}
K.K.~Sharma and R.C.~Verma, \emph{{A Study of weak mesonic decays of $\Lambda_c$ and $\Xi_c$ baryons on the basis of HQET results}}, \href{https://doi.org/10.1007/s100529801008}{\emph{Eur. Phys. J. C} {\bfseries 7} (1999) 217} [\href{https://arxiv.org/abs/hep-ph/9803302}{{\ttfamily hep-ph/9803302}}].

\bibitem{Korner:1978ec}
J.G.~K\"orner, G.~Kramer and J.~Willrodt, \emph{{Weak Decays of the Charmed Baryon C$_0^+$ and the Inclusive Yield of $\Lambda$ and $p$}}, \href{https://doi.org/10.1016/0370-2693(78)90495-1}{\emph{Phys. Lett. B} {\bfseries 78} (1978) 492}.

\bibitem{Uppal:1994pt}
T.~Uppal, R.C.~Verma and M.P.~Khanna, \emph{{Constituent quark model analysis of weak mesonic decays of charm baryons}}, \href{https://doi.org/10.1103/PhysRevD.49.3417}{\emph{Phys. Rev. D} {\bfseries 49} (1994) 3417}.

\bibitem{Lu:2016ogy}
C.-D.~L\"u, W.~Wang and F.-S.~Yu, \emph{{Test flavor SU(3) symmetry in exclusive $\Lambda_c$ decays}}, \href{https://doi.org/10.1103/PhysRevD.93.056008}{\emph{Phys. Rev. D} {\bfseries 93} (2016) 056008} [\href{https://arxiv.org/abs/arXiv:1601.04241}{{\ttfamily arXiv:1601.04241}}].

\bibitem{Savage:1989qr}
M.J.~Savage and R.P.~Springer, \emph{{SU(3) Predictions for Charmed Baryon Decays}}, \href{https://doi.org/10.1103/PhysRevD.42.1527}{\emph{Phys. Rev. D} {\bfseries 42} (1990) 1527}.

\bibitem{Kohara:1991ug}
Y.~Kohara, \emph{{Quark diagram analysis of charmed baryon decays}}, \href{https://doi.org/10.1103/PhysRevD.44.2799}{\emph{Phys. Rev. D} {\bfseries 44} (1991) 2799}.

\bibitem{Verma:1995dk}
R.C.~Verma and M.P.~Khanna, \emph{{Cabibbo favored hadronic decays of charmed baryons in flavor SU(3)}}, \href{https://doi.org/10.1103/PhysRevD.53.3723}{\emph{Phys. Rev. D} {\bfseries 53} (1996) 3723} [\href{https://arxiv.org/abs/hep-ph/9506394}{{\ttfamily hep-ph/9506394}}].

\bibitem{Chau:1995gk}
L.-L.~Chau, H.-Y.~Cheng and B.~Tseng, \emph{{Analysis of two-body decays of charmed baryons using the quark diagram scheme}}, \href{https://doi.org/10.1103/PhysRevD.54.2132}{\emph{Phys. Rev. D} {\bfseries 54} (1996) 2132} [\href{https://arxiv.org/abs/hep-ph/9508382}{{\ttfamily hep-ph/9508382}}].

\bibitem{Bigi:1994aw}
I.I.Y.~Bigi and H.~Yamamoto, \emph{{Interference between Cabibbo allowed and doubly forbidden transitions in $D \to K_{S,L} + \pi's$ decays}}, \href{https://doi.org/10.1016/0370-2693(95)00285-S}{\emph{Phys. Lett. B} {\bfseries 349} (1995) 363} [\href{https://arxiv.org/abs/hep-ph/9502238}{{\ttfamily hep-ph/9502238}}].

\bibitem{Wang:2017gxe}
D.~Wang, P.-F.~Guo, W.-H.~Long and F.-S.~Yu, \emph{{K$_{S}^{0}$-K$_{L}^{0}$ asymmetries and CP violation in charmed baryon decays into neutral kaons}}, \href{https://doi.org/10.1007/JHEP03(2018)066}{\emph{JHEP} {\bfseries 03} (2018) 066} [\href{https://arxiv.org/abs/arXiv:1709.09873}{{\ttfamily arXiv:1709.09873}}].

\bibitem{CLEO:2007rhw}
{\scshape CLEO} collaboration, \emph{{Comparison of $D \to K_S^0\pi$ and $D\to K_L^0\pi$ Decay Rates}}, \href{https://doi.org/10.1103/PhysRevLett.100.091801}{\emph{Phys. Rev. Lett.} {\bfseries 100} (2008) 091801} [\href{https://arxiv.org/abs/arXiv:0711.1463}{{\ttfamily arXiv:0711.1463}}].

\bibitem{BESIII:2022xhe}
{\scshape BESIII} collaboration, \emph{{Measurements of absolute branching fractions of $D^0\to K_L^0\phi$, $K_L^0\eta$, $K_L^0\omega$, and $K_L^0\eta^{\prime}$}}, \href{https://doi.org/10.1103/PhysRevD.105.092010}{\emph{Phys. Rev. D} {\bfseries 105} (2022) 092010} [\href{https://arxiv.org/abs/arXiv:2202.13601}{{\ttfamily arXiv:2202.13601}}].

\bibitem{BESIII:2019kfh}
{\scshape BESIII} collaboration, \emph{{Study of the Decays $D_{s}^{+} \rightarrow K_{S}^{0}K^{+}$ and $K_{L}^{0}K^{+}$}}, \href{https://doi.org/10.1103/PhysRevD.99.112005}{\emph{Phys. Rev. D} {\bfseries 99} (2019) 112005} [\href{https://arxiv.org/abs/arXiv:1903.04164}{{\ttfamily arXiv:1903.04164}}].

\bibitem{ParticleDataGroup:2022pth}
{\scshape Particle Data Group} collaboration, \emph{{Review of Particle Physics}}, \href{https://doi.org/10.1093/ptep/ptac097}{\emph{PTEP} {\bfseries 2022} (2022) 083C01}.

\bibitem{BESIII:2022dxl}
{\scshape BESIII} collaboration, \emph{{Measurement of integrated luminosities at BESIII for data samples at center-of-mass energies between 4.0 and 4.6 GeV}}, \href{https://doi.org/10.1088/1674-1137/ac80b4}{\emph{Chin. Phys. C} {\bfseries 46} (2022) 113002} [\href{https://arxiv.org/abs/arXiv:2203.03133}{{\ttfamily arXiv:2203.03133}}].

\bibitem{BESIII:2022ulv}
{\scshape BESIII} collaboration, \emph{{Luminosities and energies of e$^{+}$e$^{-}$ collision data taken between $\sqrt{s}$=4.61 GeV and 4.95 GeV at BESIII}}, \href{https://doi.org/10.1088/1674-1137/ac84cc}{\emph{Chin. Phys. C} {\bfseries 46} (2022) 113003} [\href{https://arxiv.org/abs/arXiv:2205.04809}{{\ttfamily arXiv:2205.04809}}].

\bibitem{BESIII:2009fln}
{\scshape BESIII} collaboration, \emph{{Design and Construction of the BESIII Detector}}, \href{https://doi.org/10.1016/j.nima.2009.12.050}{\emph{Nucl. Instrum. Meth. A} {\bfseries 614} (2010) 345} [\href{https://arxiv.org/abs/arXiv:0911.4960}{{\ttfamily arXiv:0911.4960}}].

\bibitem{Yu:2016cof}
C.~Yu et~al., \emph{{BEPCII Performance and Beam Dynamics Studies on Luminosity}},  in \emph{{7th International Particle Accelerator Conference}}, p.~TUYA01, 2016, \href{https://doi.org/10.18429/JACoW-IPAC2016-TUYA01}{DOI}.

\bibitem{BESIII:2020nme}
{\scshape BESIII} collaboration, \emph{{Future Physics Programme of BESIII}}, \href{https://doi.org/10.1088/1674-1137/44/4/040001}{\emph{Chin. Phys. C} {\bfseries 44} (2020) 040001} [\href{https://arxiv.org/abs/arXiv:1912.05983}{{\ttfamily arXiv:1912.05983}}].

\bibitem{Huang:2022wuo}
K.-X.~Huang, Z.-J.~Li, Z.~Qian, J.~Zhu, H.-Y.~Li, Y.-M.~Zhang et~al., \emph{{Method for detector description transformation to Unity and application in BESIII}}, \href{https://doi.org/10.1007/s41365-022-01133-8}{\emph{Nucl. Sci. Tech.} {\bfseries 33} (2022) 142} [\href{https://arxiv.org/abs/arXiv:2206.10117}{{\ttfamily arXiv:2206.10117}}].

\bibitem{Li:2017jpg}
X.~Li et~al., \emph{{Study of MRPC technology for BESIII endcap-TOF upgrade}}, \href{https://doi.org/10.1007/s41605-017-0014-2}{\emph{Radiat. Detect. Technol. Methods} {\bfseries 1} (2017) 13}.

\bibitem{Guo:2017sjt}
Y.-X.~Guo et~al., \emph{{The study of time calibration for upgraded end cap TOF of BESIII}}, \href{https://doi.org/10.1007/s41605-017-0012-4}{\emph{Radiat. Detect. Technol. Methods} {\bfseries 1} (2017) 15}.

\bibitem{Lange:2001uf}
D.J.~Lange, \emph{{The EvtGen particle decay simulation package}}, \href{https://doi.org/10.1016/S0168-9002(01)00089-4}{\emph{Nucl. Instrum. Meth. A} {\bfseries 462} (2001) 152}.

\bibitem{GEANT4:2002zbu}
{\scshape GEANT4} collaboration, \emph{{GEANT4--a simulation toolkit}}, \href{https://doi.org/10.1016/S0168-9002(03)01368-8}{\emph{Nucl. Instrum. Meth. A} {\bfseries 506} (2003) 250}.

\bibitem{YouZhengYun2008}
Y.~Zheng-Yun, L.~Yu-Tie and M.~Ya-Jun, \emph{{A method for detector description exchange among ROOT GEANT4 and GEANT3}}, \href{https://doi.org/10.1088/1674-1137/32/7/012}{\emph{Chin. Phys. C} {\bfseries 32} (2008) 572}.

\bibitem{Liang:2009zzb}
Y.-T.~Liang et~al., \emph{{A uniform geometry description for simulation, reconstruction and visualization in the BESIII experiment}}, \href{https://doi.org/10.1016/j.nima.2009.02.036}{\emph{Nucl. Instrum. Meth. A} {\bfseries 603} (2009) 325}.

\bibitem{Jadach:2000ir}
S.~Jadach, B.F.L.~Ward and Z.~Was, \emph{{Coherent exclusive exponentiation for precision Monte Carlo calculations}}, \href{https://doi.org/10.1103/PhysRevD.63.113009}{\emph{Phys. Rev. D} {\bfseries 63} (2001) 113009} [\href{https://arxiv.org/abs/hep-ph/0006359}{{\ttfamily hep-ph/0006359}}].

\bibitem{Jadach:1999vf}
S.~Jadach, B.F.L.~Ward and Z.~Was, \emph{{The Precision Monte Carlo event generator K K for two fermion final states in $e^+e^-$ collisions}}, \href{https://doi.org/10.1016/S0010-4655(00)00048-5}{\emph{Comput. Phys. Commun.} {\bfseries 130} (2000) 260} [\href{https://arxiv.org/abs/hep-ph/9912214}{{\ttfamily hep-ph/9912214}}].

\bibitem{Ping:2008zz}
R.-G.~Ping, \emph{{Event generators at BESIII}}, \href{https://doi.org/10.1088/1674-1137/32/8/001}{\emph{Chin. Phys. C} {\bfseries 32} (2008) 599}.

\bibitem{Chen:2000tv}
J.C.~Chen, G.S.~Huang, X.R.~Qi, D.H.~Zhang and Y.S.~Zhu, \emph{{Event generator for $J/\psi$ and $\psi(2S)$ decay}}, \href{https://doi.org/10.1103/PhysRevD.62.034003}{\emph{Phys. Rev. D} {\bfseries 62} (2000) 034003}.

\bibitem{Richter-Was:1992hxq}
E.~Richter-Was, \emph{{QED bremsstrahlung in semileptonic B and leptonic tau decays}}, \href{https://doi.org/10.1016/0370-2693(93)90062-M}{\emph{Phys. Lett. B} {\bfseries 303} (1993) 163}.

\bibitem{BESIII:2017kqg}
{\scshape BESIII} collaboration, \emph{{Precision measurement of the $e^{+}e^{-}~\rightarrow~\Lambda_{c}^{+} \bar{\Lambda}_{c}^{-}$ cross section near threshold}}, \href{https://doi.org/10.1103/PhysRevLett.120.132001}{\emph{Phys. Rev. Lett.} {\bfseries 120} (2018) 132001} [\href{https://arxiv.org/abs/arXiv:1710.00150}{{\ttfamily arXiv:1710.00150}}].

\bibitem{BESIII:2023rwv}
{\scshape BESIII} collaboration, \emph{{Measurement of Energy-Dependent Pair-Production Cross Section and Electromagnetic Form Factors of a Charmed Baryon}}, \href{https://doi.org/10.1103/PhysRevLett.131.191901}{\emph{Phys. Rev. Lett.} {\bfseries 131} (2023) 191901} [\href{https://arxiv.org/abs/arXiv:2307.07316}{{\ttfamily arXiv:2307.07316}}].

\bibitem{BESIII:2019odb}
{\scshape BESIII} collaboration, \emph{{Measurements of Weak Decay Asymmetries of $\Lambda_c^+\to pK_S^0$, $\Lambda\pi^+$, $\Sigma^+\pi^0$, and $\Sigma^0\pi^+$}}, \href{https://doi.org/10.1103/PhysRevD.100.072004}{\emph{Phys. Rev. D} {\bfseries 100} (2019) 072004} [\href{https://arxiv.org/abs/arXiv:1905.04707}{{\ttfamily arXiv:1905.04707}}].

\bibitem{MARK-III:1985hbd}
{\scshape MARK-III} collaboration, \emph{{Direct Measurements of Charmed D Meson Hadronic Branching Fractions}}, \href{https://doi.org/10.1103/PhysRevLett.56.2140}{\emph{Phys. Rev. Lett.} {\bfseries 56} (1986) 2140}.

\bibitem{MARK-III:1987jsm}
{\scshape MARK-III} collaboration, \emph{{A Reanalysis of Charmed D Meson Branching Fractions}}, .

\bibitem{Ke:2023qzc}
B.-C.~Ke, J.~Koponen, H.-B.~Li and Y.~Zheng, \emph{{Recent Progress in Leptonic and Semileptonic Decays of Charmed Hadrons}}, \href{https://doi.org/10.1146/annurev-nucl-110222-044046}{\emph{Ann. Rev. Nucl. Part. Sci.} {\bfseries 73} (2023) 285} [\href{https://arxiv.org/abs/arXiv:2310.05228}{{\ttfamily arXiv:2310.05228}}].

\bibitem{Li:2021iwf}
H.-B.~Li and X.-R.~Lyu, \emph{{Study of the standard model with weak decays of charmed hadrons at BESIII}}, \href{https://doi.org/10.1093/nsr/nwab181}{\emph{Natl. Sci. Rev.} {\bfseries 8} (2021) nwab181} [\href{https://arxiv.org/abs/2103.00908}{{\ttfamily 2103.00908}}].

\bibitem{BESIII:2022xne}
{\scshape BESIII} collaboration, \emph{{Observations of the Cabibbo-Suppressed decays $\Lambda_{c}^{+}\to n\pi^{+}\pi^{0}$, $n\pi^{+}\pi^{-}\pi^{+}$ and the Cabibbo-Favored decay $\Lambda_{c}^{+}\to nK^{-}\pi^{+}\pi^{+}$}}, \href{https://doi.org/10.1088/1674-1137/ac9d29}{\emph{Chin. Phys. C} {\bfseries 47} (2023) 023001} [\href{https://arxiv.org/abs/arXiv:2210.03375}{{\ttfamily arXiv:2210.03375}}].

\bibitem{BESIII:2015bjk}
{\scshape BESIII} collaboration, \emph{{Measurements of absolute hadronic branching fractions of $\Lambda_{c}^{+}$ baryon}}, \href{https://doi.org/10.1103/PhysRevLett.116.052001}{\emph{Phys. Rev. Lett.} {\bfseries 116} (2016) 052001} [\href{https://arxiv.org/abs/arXiv:1511.08380}{{\ttfamily arXiv:1511.08380}}].

\bibitem{BESIII:2022bkj}
{\scshape BESIII} collaboration, \emph{{Observation of the Singly Cabibbo Suppressed Decay $\Lambda^+_c \to n \pi^+$}}, \href{https://doi.org/10.1103/PhysRevLett.128.142001}{\emph{Phys. Rev. Lett.} {\bfseries 128} (2022) 142001} [\href{https://arxiv.org/abs/arXiv:2201.02056}{{\ttfamily arXiv:2201.02056}}].

\bibitem{BESIII:2012mpj}
{\scshape BESIII} collaboration, \emph{{Search for hadronic transition $\chi_{cj} \to \eta_c\pi^+\pi^-$ and observation of $\chi_{cj} \to K\overline{K}\pi\pi\pi$}}, \href{https://doi.org/10.1103/PhysRevD.87.012002}{\emph{Phys. Rev. D} {\bfseries 87} (2013) 012002} [\href{https://arxiv.org/abs/arXiv:1208.4805}{{\ttfamily arXiv:1208.4805}}].

\bibitem{BESIII:2022ahw}
{\scshape BESIII} collaboration, \emph{{Observation of the $J/\psi$ and $\psi(3686)$ decays into $\eta\Sigma^+\Sigma^-$}}, \href{https://doi.org/10.1103/PhysRevD.106.112007}{\emph{Phys. Rev. D} {\bfseries 106} (2022) 112007} [\href{https://arxiv.org/abs/arXiv:2210.09601}{{\ttfamily arXiv:2210.09601}}].

\end{thebibliography}\endgroup
\newpage

\newpage
\section*{The BESIII collaboration}
\addcontentsline{toc}{section}{The BESIII collaboration}

\begin{small}
    \begin{center}
    M.~Ablikim$^{1}$, M.~N.~Achasov$^{4,c}$, P.~Adlarson$^{76}$, O.~Afedulidis$^{3}$, X.~C.~Ai$^{81}$, R.~Aliberti$^{35}$, A.~Amoroso$^{75A,75C}$, Q.~An$^{72,58,a}$, Y.~Bai$^{57}$, O.~Bakina$^{36}$
% ...
\\
\vspace{0.2cm}
(BESIII Collaboration)\\
\vspace{0.2cm} {\it
$^{1}$ Institute of High Energy Physics, Beijing 100049, People's Republic of China\\
% ...
$^{3}$ Bochum  Ruhr-University, D-44780 Bochum, Germany\\
$^{4}$ Budker Institute of Nuclear Physics SB RAS (BINP), Novosibirsk 630090, Russia\\
% ...
$^{35}$ Johannes Gutenberg University of Mainz, Johann-Joachim-Becher-Weg 45, D-55099 Mainz, Germany\\
$^{36}$ Joint Institute for Nuclear Research, 141980 Dubna, Moscow region, Russia\\
% ...
$^{57}$ Southeast University, Nanjing 211100, People's Republic of China\\
$^{58}$ State Key Laboratory of Particle Detection and Electronics, Beijing 100049, Hefei 230026, People's Republic of China\\
% ...
$^{72}$ University of Science and Technology of China, Hefei 230026, People's Republic of China\\
% ...
$^{75}$ University of Turin and INFN, (A)University of Turin, I-10125, Turin, Italy; (B)University of Eastern Piedmont, I-15121, Alessandria, Italy; (C)INFN, I-10125, Turin, Italy\\
$^{76}$ Uppsala University, Box 516, SE-75120 Uppsala, Sweden\\
% ...
$^{81}$ Zhengzhou University, Zhengzhou 450001, People's Republic of China\\

\vspace{0.2cm}
$^{a}$ Deceased\\
% ...
$^{c}$ Also at the Novosibirsk State University, Novosibirsk, 630090, Russia\\
% ...
    }\end{center}
    
    \vspace{0.4cm}
    \end{small}
\end{document}